\documentclass[rnote]{aa}
\usepackage[varg]{txfonts}

\usepackage{natbib}
\bibpunct{(}{)}{;}{a}{}{,} 
\usepackage{graphicx}
\usepackage{color}
\usepackage[colorlinks,citecolor=blue]{hyperref}
\usepackage{amstext}

\begin{document}

\title{Parameterization of the level-resolved radiative recombination rate coefficients for the SPEX code}

\author{Junjie Mao\inst{\ref{inst1},~\ref{inst2}}, 
        \and Jelle Kaastra\inst{\ref{inst1},~\ref{inst2}}}

\offprints{J. Mao,~\email{J.Mao@sron.nl}}

\institute{SRON Netherlands Institute for Space Research, Sorbonnelaan 2, 3584 CA Utrecht,  
           the Netherlands \label{inst1}
           \and Leiden Observatory, Leiden University, Niels Bohrweg 2, 2300 RA Leiden, the   
              Netherlands \label{inst2}}

\date{Received date / Accepted date}

\abstract{
The level-resolved radiative recombination (RR) rate coefficients for H-like to Na-like ions
from H ($Z=1$) up to and including Zn ($Z=30$) are studied here. For H-like ions,
the quantum-mechanical exact photoionization cross sections for nonrelativistic
hydrogenic systems are usedto calculate the RR rate coefficients 
under the principle of detailed balance, while for He-like to Na-like ions, 
the archival data on ADAS are adopted. 
Parameterizations are made for the direct capture rates 
in a wide temperature range.
The fitting accuracies are better than $5\%$ for about $99\%$ of 
the $\sim3\times10^4$ levels considered here. 
The $\sim1\%$ exceptions include
levels from low-charged many-electron ions, 
and/or high-shell ($n \gtrsim 4$) levels are less important
in terms of interpreting X-ray emitting astrophysical plasmas.
The RR data will be incorporated into 
the high-resolution spectral analysis package SPEX.
}

\titlerunning{Parameterization of level-resolved RR data for SPEX}
\authorrunning{Mao \& Kaastra}
\maketitle

\section{Introduction}
\label{sct:intro}
Some astrophysical plasmas, for instance, the intracluster medium (ICM), 
are generally not in local temperature equilibrium (LTE).
To determine the ionization state of these plasmas,
the individual collisional and radiative ionization and recombination processes 
need to be considered in great detail. 
The recombination processes can be divided into three subclasses: 
(resonant) dielectronic recombination (DR), 
(nonresonant) radiative recombination (RR),
and charge-exchange recombination (CXR). Generally speaking, 
DR is the dominant recombination process 
for hot plasmas compared with RR for most ions \citep{bry09}. On the other hand, 
when the temperature of the plasma is low enough 
for  neutral atoms, molecules, 
and ions to co-exist, CXR can be a process that competes with DR \citep{arn85}.  
Nevertheless, RR cannot be ignored at least in some temperature ranges
in terms of the total recombination rate.
In addition, knowledge of the level-resolved RR rate coefficient 
is required to calculate the emission line spectrum of astrophysical plasmas, 
for instance, the collisional ionization equilibrium (CIE) spectrum
and the radiative recombination spectrum \citep{tuc66}. 

In Sect.~\ref{sct:bkgd} we briefly summarize the main points of the previous
studies on RR data, focusing on the parameterizations for the total RR rates. 
In Sect.~\ref{sct:hlike} we show the details of the numerical approaches 
to derive the RR rate coefficients for H-like ions. 
Subsequently, in Sect.~\ref{sct:he2nalike}, 
we describe to which extent the archival data on ADAS are analyzed
for He-like to Na-like ions.
Details of the fitting strategy are shown in Sect.~\ref{sct:fit}.
Results of the parameterizations, available in CDS, 
are discussed in Sect.~\ref{sct:rnd}. 

\section{Historical background}
\label{sct:bkgd}
Key issues for atomic data are always how precisely 
the analytical and/or numerical calculations can be made, 
and how to parameterize the data for a full description, 
yet in a quick and accurate way. 

The SPEX \citep{kaa96} code, aiming at interpreting 
high-energy resolution X-ray spectra, 
allows users to make fast (all the calculations are run on the fly)
yet adequately accurate analyses, including spectral fitting,
plotting, and diagnostic output of the astrophysical plasma.
Driven by the practical user experience, 
the number of mathematical operations used and atomic data storage
for the complete description of the basic atomic processes 
need to be minimized. 
The SPEX code makes great efforts to 
parameterize the detailed atomic data as
best possible.
The parameterization of the level-resolved RR rate coefficients
is one of them.

Previously, only the total RR rates were parameterized, 
and we briefly summarize them here.   
To begin with, pioneering work was done by \citet{sea59}.
With the asymptotic expansion of the Gaunt factor, 
the photoionization cross sections (PICSs) of the hydrogenic ions were derived. 
And then the total RR rates are fitted with
\begin{eqnarray}
\small
\alpha^{\rm RR}_{\rm ttl}(T) &=& 5.197\times10^{-14}~Z~\lambda^{-1/2}~(0.4288 + 0.5~\ln \lambda \nonumber \\
&& + 0.469~\lambda^{-1/3})~{\rm cm^{3}~s^{-1}}~,\end{eqnarray}
where $\lambda = Z^2 \frac{E_{\rm H}}{kT}$, $Z$ is the atomic number, 
and $E_{\rm H}$ is the Rydberg unit of energy.

\citet{ald73} calculated the radiative recombination data for the non-hydrogenic ions of 
He, C, N, O, Ne, Mg, Si, and S by using the ground-state PICSs from literature 
and extrapolating along the isoelectronic sequences. Furthermore, a power-law (PL) 
fitting function for the total RR rates was proposed, 
\begin{equation}
\small
\alpha^{\rm RR}_{\rm ttl}(T) = A~T_4^{-\eta}~,
\label{eq:fit_ald73}
\end{equation}
where $A$ and $\eta$ are the fitting parameters, 
and $T_4$ is the electron temperature in units of $10^4$~K. 
The PL model was subsequently also favored by 
\citet{shu82}, \citet{arn85}, \citet{lan90} and \citet{lan91}. 

\citet{peq91} took advantage of the latest references available at that time 
for the ground states and part of the excited states ($n=2$~or $3$) of PICSs, 
as well as the proper extrapolation and scaling law (analogy with the hydrogenic ions). 
RR data for H, He, C, N, O, and Ne ions were obtained. 
The total RR rates were fit to the following expression: 
\begin{equation}
\small
\alpha^{\rm RR}_{\rm ttl}(T) = 10^{-13}~Z~\frac{a~t^b}{1 + c~t^d}~{\rm cm^{3}~s^{-1}}~,
\label{eq:fit_peq91}
\end{equation} 
where $t = {T_4}/{Z^2}$ and ${Z}$ is the ionic charge ($Z=1$ for recombination toward 
the neutral state). 

Using the Hartree-Fock wave functions, the standard partial PICSs 
for subshells ($nl$) up to $5g$ of He-like through Al-like ions 
were provided by \citet{cla86}. Based on this important progress, 
\citet{arn92} calculated the RR data of ${\rm Fe}^{+14}$ through ${\rm Fe}^{+25}$.
The total RR rates of these iron ions in the temperature range of $10^5-10^8$~K 
were described (with a fitting accuracy better than 5\%) by a log parabola function
\begin{equation}
\small
\alpha^{\rm RR}_{\rm ttl}(T) = A~T_4^{-\alpha - \beta~\log T_4}~,
\label{eq:fit_arn92}
\end{equation} 
where $A,~\alpha$ and $\beta$ are the fitting parameters. 
Additionally, \citet{maz98} summarized the RR data mentioned above,
and more importantly, updated the ionization balance for optically thin plasmas
by collecting all the available ionization and recombination (RR + DR) data.

\citet{ver96a} extended the calculation of the RR data for the radiative recombination 
toward H-like, He-like, Li-like, and Na-like ions for all the elements 
from H up to and including Zn. 
For the PICSs of the ground states of He-like, Li-like, and Na-like ions, 
the latest results provided by \citet{ver96b} were adopted. 
For those of the excited states with $n\le5$ of the highly ionized 
(at least five times) species, the partial PICSs of \citet{cla86} were used.
To calculate excited states with $n\le 10$ of the first four ionization states, 
correction for incomplete shielding \citep{gou78} was required. 
The hydrogenic approximation was used for the remaining states. 
We note that a comparison with the Opacity Project (OP) calculations \citep{sea92} were made, 
and the conclusion was that the accuracies were better than 10\%.  
The fitting model used for the total RR rates 
in the temperature range from 3~K to $10^{10}$~K is 
\begin{equation}
\small
\alpha^{\rm RR}_{\rm ttl}(T) = a \left[\sqrt{\frac{T}{T_0}}~\left(1 + \sqrt{\frac{T}{T_0}}\right)^{1 - b}~\left(1 + \sqrt{\frac{T}{T_1}}\right)^{1 + b}\right]^{-1}~,
\label{eq:fit_ver96}
\end{equation}
where $a,~b,~T_0$, and $T_1$ are the fitting parameters.

\citet{gu03} used a distorted-wave approximation, similar to the Dirac-Hartree-Slater model
used in \citet{ver93}, to calculate the detailed PICSs with $n \le 10$ for 
bare through F-like isoelectronic sequences of Mg, Si, S, Ar, Ca, Fe, and Ni.
For $n > 10$ shells the semiclassical Kramers formula was used. 
The computational procedure of the flexible atomic code \citep[FAC, ][]{gu03}, 
which provides not only the RR data, but also other important atomic data, 
is similar to the fully relativistic distorted-wave method of \citet{zha98}.
Based on the fitting model given by \citet{ver96a}, \citet{gu03} used a modified expression 
to fit the total RR rates in the temperature range from $10^{-4}$ to $10^4$~eV,
that is, $10$~K to $10^8$~K.  
Compared to Eq.~(\ref{eq:fit_ver96}), one necessary modification is that 
for some elements (e.g., F-like Mg, O-like Mg, F-like Si) parameter $b$ is replaced by 
$b + b_1~\exp{\left(T_2/T\right)}$, 
which means that two more parameters $b_1$ and $T_2$ are required. 
The fitting accuracies are within 5\% for the entire temperature range. 

\citet{bad06} used the AUTOSTRUCTURE code \citep{bad86, bad03} to calculate PICSs, 
thus the RR data, for all the elements up to and including Zn, 
plus Kr, Mo, and Xe, for all the isoelectronic sequences up to Na-like forming Mg-like 
in the temperature range of $Z^2(10^1-10^7)$~K, where $Z$ is the atomic number. 
Similar to \citet{gu03}, \citet{bad06} adopted Eq.~(\ref{eq:fit_ver96}) 
to fit the total RR rates. 
Likewise, the additional fitting parameters $b_1 \text{and}~T_2$ 
are also required for some of the low-charge ions because of the highly nonhydrogenic screening of the wave functions for 
the low--${nl}$ states in low-charged many-electron ions. 
The fits are accurate to within 5\% for singly and doubly ionized ions, 
and better than 1\% for multiply charged ions. 

In addition, Nahar and coworkers, for example, \citet{nah99}, obtained
the total (unified DR + RR) rate coefficients for various ions
with their \textit{R}-matrix calculations. Meanwhile,
adopting most recent RR and DR calculations,
\citet{bry06} updated the collisional ionization equilibrium 
for optically thin plasmas.

Throughout the entire analysis, we refer to the recombined ion
when we speak of the radiative recombination of a certain ion.

\section{RR rate coefficients for H-like ions}
\label{sct:hlike}
\subsection{Photoionization cross sections}
\label{sct:pics}
For the hydrogen sequence, the archival data on ADAS  
are ready to use, with fully $nLSJ$ resolved RR rate coefficients 
complete up to n = 8. 
With the calculation details described below, 
we completed the rate coefficients for all the levels up to n=16. 

The time-reversed process of (dielectronic and radiative) recombination is
(resonant and non-resonant) photoionization (PI). 
Therefore, in turn, 
radiative recombination cross sections (RRCSs) 
can be obtained through the Milne relation under the principle of detailed balance 
(or microscopic reversibility) for PICSs.  
The exact PICS for nonrelativistic hydrogenic systems
can be obtained with the quantum mechanical treatment provided
by \citet[][SH91]{sto91}.
Their FORTRAN code, based on recursion techniques,
yields accurate, stable, and fast numerical evaluations of
bound-free PICSs (and bound-bound electron dipole transition probability)
for nonrelativistic hydrogenic systems with $n$ up to 500.
Apparently, PICSs of hydrogenic systems can also be precisely calculated
with the AUTOSTRUCTURE \citep{bad86} code.

When PICSs ($\sigma^{\rm PI}_{n(l)}$) are available, 
RRCSs ($\sigma^{\rm RR}_{n(l)}$, which refers to recombination to the $n$th shell 
or to the subshell $nl$, respectively) 
can be obtained through the Milne relation under the principle of 
detailed balance.

Alternatively, if only the total RR rates $(\alpha_{\rm ttl}^{\rm RR})$ are needed, 
since the numerical Maxwellian convolution of the RR rates 
to $n \ge 100$ shells are computationally expensive, 
the semiclassical Kramers formula of the RRCS 
can be used instead to save computational time, 
similar to the approach presented by \citet{gu03}.

\subsection{Radiative recombination data}
\label{sct:rr_data}
The ${n(l)}$-resolved radiative recombination rate coefficients $R_{n(l)}$ 
can then be calculated by 
\begin{equation}
\small
R_{n(l)} = \int_0^{\infty} v_e f(v_e) \sigma^{\rm RR}_{n(l)}(v_e) dv_e~{\rm cm^{3}~s^{-1}}~,
\label{eq:rrc}
\end{equation}
where $f(v_e)$ is the probability density distribution of the velocity of the free electrons, 
and the Maxwell-Boltzmann distribution for the free electrons is adopted in the following calculation. 
Accordingly, the total radiative recombination rates 
\begin{equation}
\small
\alpha^{\rm RR}_{\rm ttl} =  \sum_n R_{n}~.
\label{eq:rrr}
\end{equation}

We note that the level-resolved RR rate coefficients can be obtained through
the term-resolved RR rate coefficients
\begin{equation}
\small
R_{\rm lev} = \frac{(2J + 1)}{(2S + 1) (2L + 1)} R_{nl}~,
\label{eq:lsj}
\end{equation}
where $L$ is the angular momentum quantum number,
$S$ is the spin quantum number, and for H-like ions $S=1/2$,
and $J$ is the total angular momentum quantum number.
This distribution (Eq.~\ref{eq:lsj})
agrees with the ADAS term-resolved (LS coupling) 
and level-resolved (intermediate coupling) RR rate coefficients 
of the hydrogenic systems with $n<=8$.

\section{RR rate coefficients for He-like to Na-like ions}
\label{sct:he2nalike}
For RR rate coefficients of He-like to Na-like ions, 
the archival RR data of ADAS were used. 
We note that there are two sets of RR data for each ion: 
the one calculated with LS coupling is term-resolved (or ${nLS}$ resolved), 
while the other, calculated with intermediate coupling, is level-resolved
(or ${nLSJ}$ resolved). Only the level-resolved data are analyzed here. 
Moreover, the ADAS RR data (both LS and intermediate coupling) 
cover not only radiative recombinations from 
the ground state of the recombining ion, but also from meta-stable states.
Even more complicated, the ground state is not necessary identical to the ground level.
For C-like to F-like ions, 
there are fine-structure levels within the ground term of the recombining ion. 
For F-like ions, for example, the ground term is ($1s^2 2s^2 2p^4,~^3P$), and
accordingly, the fine-structure levels are $^3P_2$ (the ground level),
$^3P_0$ and $^3P_1$.
The RR rate coefficients from the excited fine-structure levels 
are lower than those from the ground level as a result of the additional auto-ionization pathways \citep{bad06}. 
We here only carried out the parameterization 
for RR from the ground level of the recombining ion.

Additionally, all the levels were coded according to intrinsic level indices of SPEX. 
The advantage is that a set of group numbers
are included to distinguish (by different group numbers) 
those levels with the exact same $nLSJ$ quantum numbers and 
slightly different configuration for many-electron ions. 
For instance, for O-like ions, the electron configuration of 
$1s^2 2s^2 2p^3 (^2P)~np$ (where $n \ge$3) has a $^1P$ term, 
where $^2P$ in the parenthesis denotes 
the coupling of the $2p^3$ electron configuration,  
but also the electron configuration of $1s^2 2s^2 2p^3 (^2D) np$.

\section{Fitting strategy}
\label{sct:fit}
For the hydrogenic systems, the RR data are calculated (following SH91) 
in a wide temperature grid ranging from $10^1$ to $10^8$~K,
with ten steps per decade on a $\log_{10}$-scale. 
At even higher temperature $T > 10^8$~K, in principle, 
relativistic effects for the large-$Z$ elements should be taken into account. 
However, differences that are due to relativistic effects
may not play an important role because the RR rate coefficients
at high temperature are lower than those at low temperature, and more importantly, 
lower than the rate coefficients of other processes at high temperature. 
Therefore, when the RR data for $T > 10^8$~K are not calculated,
extrapolation to $T > 10^8$~K should be feasible. 
For the He-like to Na-like ions, the temperature sets in 
$z^2 (10^1-10^7)$~K on ADAS were used, 
where $z$ is the nuclear charge of the recombining ion.  

Sherpa\footnote{http://cxc.harvard.edu/sherpa/index.html} 
was used for the fitting procedure, with its Simplex (i.e., Neldermead)
optimization method. 
We propose the following model function for the fitting:
\begin{equation}
\small
R(T) = 10^{-10}~{\rm cm^3~s^{-1}}~a_{0}~T^{-b_{0} - c_{0}\ln{T}}~\left(\frac{1 + a_2 T^{-b2}}{1 + a_1 T^{-b1}}\right)~,
\label{eq:fit_mao15}
\end{equation}
where the electron temperature $T$ is in units of eV, 
$a_{0,~1}~{\rm and}~b_{0,~1}$ are primary fitting parameters, 
$c_0$ and $(a_2,~b_2)$ are additional fitting parameters. 
The additional parameters were frozen to zero if they were not used.
Furthermore, we constrained $b_{0-2}$ to between -10.0 and 10.0 
and $c_0$ to between 0.0 and 1.0. 
The initial values of the four primary fitting parameters 
$a_{0,~1}~{\rm and}~b_{0,~1}$ were set to unity together with the two additional fitting parameters
$a_2~{\rm and}~b_2$ when these were thawed.
Conversely, the initial value of $c_0$, if it was thawed,
was set to either side of its boundary, 
that is, $c_0 = 0.0$ or $c_0 = 1.0$ (both were performed).

To estimate the goodness of fit,
the fits were performed with a set of artificial relative errors $(r)$.
We started with $r=1.25\%$, followed by increasing $r$ by a factor of two,
up to and including $5.0\%$. 
The chi-squared statistics adopted here were   
\begin{equation}
\small
\chi^2 = \sum_{i = 1}^{N} \left(\frac{n_i - m_i}{r~n_i}\right)^2~, 
\label{eq:chi2}
\end{equation}
where $n_i$ is the $i$th numerical calculation result and $m_i$ is the $i$th
model prediction (Eq.~\ref{eq:fit_mao15}).

On the other hand, for the model comparison, 
the RR data were first fit with the simplest model 
(i.e., the three additional parameters were frozen to zero), 
followed by attempted fits with free $c_0$ and/or free $(a_2, ~b_2)$. 
We used the chi-squared distribution for all the fitting statistics, 
and thawing one additional parameter decreases the degrees of freedom by one; 
therefore, only if the final statistic ($\chi^2$) improves 
by at least 2.71, 4.61, or 6.25 for one, two, or three 
additional free parameter(s) (at a 90\% confidence level) is
the more complicated model favored.

From these attempted fits and
considering both the fitting accuracy and simplicity of the model,
we determined the best-fit model.

\section{Results and discussion}
\label{sct:rnd}
\subsection{Total RR rates $\alpha_{\rm ttl}^{\rm RR}$}
\label{sct:fit_alpha}
The total RR rates, adding up contributions from $n=1$ up to $n = 10^4$, 
for all the H-like ions considered here 
can be parameterized with Eq.~(\ref{eq:fit_mao15}). 
This also holds for the total RR rates for He-like to Na-like ions, but no significant improvement can be achieved
compared to previous results from \citet{bad06}.

\subsection{Level-resolved RR rate coefficients}
\label{sct:fit_lev}
The fitting results, available at the CDS, contain the following information: 
Column 1 lists the isoelectronic sequence number of the recombined ion,
Col. 2 gives the atomic number, Cols. 3-9 list the fitting parameters, 
Cols. 10-12 are the degree of freedom (d.o.f.), the final statistics ($\chi^2$),
and the maximum deviation ($\delta_{\rm max}$). 
Columns 13-16 indicate the $nLSJ$ quantum numbers,
and Col. 17 lists the electron configuration.
The fitting accuracies are generally better than $5\%$ 
with the exceptions described below.

The most inaccurate fitting result ($\delta_{\rm max} = 5.8\%$) for the helium sequence
captures the free electron to the $(1s~6p,~^3 P_0)$ level
to form He-like copper (Cu XXVIII).
As for the lithium sequence,
the poorest fitting ($\delta_{\rm max} = 7.2\%$) is achieved
for the $(1s^2 6s,~^2 S_{1/2})$ level of Li-like chromium (Cr XXII).
In the beryllium sequence, the $(1s^2 2s~8s,~^1 S_0)$ level
of the beryllium atom (Be I) deviates most strongly
with $7.8\%$.
From the beryllium sequence,
the characteristic high-temperature bump
for the low-charged many-electron ions is present.
In the boron sequence, the poorest fitting is the $(1s^2 2s^2 8s,~^2 S_{1/2})$ level
of the boron atom (B I), with $\delta_{\rm max}=12.2\%$.
However, the RR to this level is
overwhelmingly dominated by other channels within the same $n=8$ shell,
except for the high-temperature end.
Nevertheless, at the high-temperature end,
the high shell RR rate coefficients are merely a few percent
lower than those of the low shell. In this case, for instance,
at $T=1$~keV (or $10^7$~K),
the rate coefficient to the $n=8$ shell is
lower by a few percent than that of the $n=2$ shell.
Similarly, in the carbon and nitrogen sequences, 
all the poorly fitted ($\delta_{\rm max} \gtrsim 10\%$) levels
of the carbon atom (C I), nitrogen atom (N I)
are also dominated by other channels within their corresponding shells.
Again, similar arguments hold for the oxygen sequence, 
not only for the poorly fitted high shell levels of the oxygen atom (O I), 
but also for the poorly fitted high shell ($n\gtrsim 4$) levels 
from the low-charged many-electron ions of O-like fluorine (F II)
and O-like neon (Ne III).
As expected, similar results are found in the fluorine sequence
for the fluorine atom (F I)
and F-like neon (Ne II),  in the neon sequence, the neon atom (Ne I)
and Ne-like soldium (Na II), in the sodium sequence, 
the soldium atom (Na I) and Na-like magnesium (Mg II).

 The relative ion fractions of all the neutral atoms are negligible
($\lesssim0.005$) for a CIE plasma with a plasma temperature as low as 5~eV and assuming the ionization balance of \citet{bry09}; we highlighted the $\sim 1\%$ poor fits above.
While the ion fraction of the low-charged many-electron ions can be $\sim 0.6$
in this extreme case, only a few tens of the high shell levels
(in total for F II, Ne II, Na II, Mg II, Ne III, etc.) are poorly fitted.
In general, a CIE plasma, such as the intracluster medium (ICM) with 
a temperature of $\sim 1$~keV, does not suffer from the poor fits at all.
The situation can be difficult for a non-equilibrium ionization (NEI) plasma
or a photoionizied plasma, while emission coming from these poorly fitted levels
is either too weak or entirely absent in the X-ray wavelength range.
Introducing more parameters might improve the fitting results, 
but for simplicity, we did not add more parameters.
Alternatively, if emission from the full wavelength range is needed
or very high accuracy is required, interpolation with the original ADAS data should be adopted.

\section{Summary}
\label{sct:sum}
We parameterized for the first time the level-resolved 
radiative recombination rate coefficients for H-like to Na-like ions
from hydrogen up to and including zinc ($Z=30$)
in a wide temperature range.
For the the hydrogen sequence, we calculated the RR data 
with the photoionization cross sections for 
nonrelativistic hydrogenic systems provided by \citet{sto91}.
The fully $nLSJ$-resolved levels are complete up to $n=16$ for H-like ions.
For helium to sodium sequences, the archival data from ADAS \citep{bad06}
were adopted, with levels complete up to $n=8$.
The bulk ($\sim99\%$) of the $3\times 10^4$ levels are fitted 
with accuracies better than 5\%.
The $\sim 1\%$ exceptions that yield relative poor fitting accuracies from 5\% to 40\%
are less important in terms of interpreting X-ray emitting astrophysical plasmas.

Together with updated inner shell ionization data 
(I. Urdampilleta et al. in prep.), 
a charge-exchange model \citep[][accepted]{gu16} 
and other atomic data will be included in the upcoming version (3.0) of 
the high-resolution X-ray spectral modeling and fitting code SPEX \citep{kaa96}, 
which will be highly useful once Astro-H/SXS data become available.

\begin{acknowledgements}
J.J.M. acknowledges discussions and consultations 
with L. Gu and A. J. J. Raassen.
SRON is supported financially by NWO, 
the Netherlands Organization for Scientific Research.
\end{acknowledgements}


\begin{thebibliography}{}

\bibitem[Aldrovandi \& Pequignot(1973)]{ald73} Aldrovandi, S.~M.~V., \& Pequignot, D.\ 1973,
\aap, 25, 137

\bibitem[Arnaud \& Rothenflug(1985)]{arn85} Arnaud, M., \& Rothenflug, R.\ 1985, \aaps, 60, 425

\bibitem[Arnaud \& Raymond(1992)]{arn92} Arnaud, M., \& Raymond, J.\ 1992, \apj, 398, 394

\bibitem[Badnell(1986)]{bad86} Badnell, N.~R.\ 1986, Journal of Physics B Atomic Molecular Physics,
19, 3827

\bibitem[Badnell \& Seaton(2003)]{bad03} Badnell, N.~R., \& Seaton, M.~J.\ 2003,
Journal of Physics B Atomic Molecular Physics, 36, 4367

\bibitem[Badnell(2006)]{bad06} Badnell, N.~R.\ 2006, \apjs,
167, 334

\bibitem[Bryans et al.(2006)]{bry06} Bryans, P., Badnell, N.~R., Gorczyca,
T.~W., et al.\ 2006, \apjs, 167, 343

\bibitem[Bryans et al.(2009)]{bry09} Bryans, P., Landi, E., \& Savin, D.~W.\ 2009, \apj, 691, 1540

\bibitem[Burgess \& Summers(1976)]{bur76} Burgess, A., \& Summers, H.~P.\ 1976, \mnras, 174, 345

\bibitem[Clark et al.(1986)]{cla86} Clark, R.~E.~H., Cowan,
R.~D., \& Bobrowicz, F.~W.\ 1986, Atomic Data and Nuclear Data Tables, 34, 415

\bibitem[Cunto et al.(1993)]{cun93} Cunto, W., Mendoza, C., Ochsenbein, F., \& Zeippen, C.~J.\ 1993,
\aap, 275, L5

\bibitem[Gould(1978)]{gou78} Gould, R.~J.\ 1978, \apj, 219, 250

\bibitem[Gu(2003)]{gu03} Gu, M.~F.\ 2003, \apj, 589, 1085

\bibitem[Gu et al. (2016)]{gu16} Gu, L., Kaastra, J., \& Raassen, A., \eprint[arXiv]{1601.05958}

\bibitem[Landini \& Monsignori Fossi(1990)]{lan90} Landini, M., \& Monsignori Fossi, B.~C.\ 1990, \aaps, 82, 229

\bibitem[Landini \& Fossi(1991)]{lan91} Landini, M., \& Fossi, B.~C.~M.\ 1991, \aaps, 91, 183

\bibitem[Kaastra et al.(1996)]{kaa96} Kaastra, J.~S., Mewe, R.,
\& Nieuwenhuijzen, H.\ 1996, UV and X-ray Spectroscopy of Astrophysical and Laboratory Plasmas, 411

\bibitem[Kaastra et al.(2008)]{kaa08} Kaastra, J.~S., Paerels, F.~B.~S., Durret,
F., Schindler, S., \& Richter, P.\ 2008, \ssr, 134, 155

\bibitem[Mazzotta et al.(1998)]{maz98} Mazzotta, P., Mazzitelli, G., Colafrancesco, S.,
\& Vittorio, N.\ 1998, \aaps, 133, 403

\bibitem[Nahar(1999)]{nah99} Nahar, S.~N.\ 1999, \apjs, 120, 131

\bibitem[Pequignot et al.(1991)]{peq91} Pequignot, D., Petitjean, P., \& Boisson, C.\ 1991,
\aap, 251, 680

\bibitem[Tucker \& Gould(1966)]{tuc66} Tucker, W.~H., \& Gould, R.~J.\ 1966, \apj, 144, 244 

\bibitem[Seaton(1959)]{sea59} Seaton, M.~J.\ 1959, \mnras, 119, 81

\bibitem[Seaton et al.(1992)]{sea92} Seaton, M.~J., Zeippen, C.~J., Tully, J.~A., et al.
\ 1992, \rmxaa, 23, 19

\bibitem[Shull \& van Steenberg(1982)]{shu82} Shull, J.~M., \& van Steenberg, M.\ 1982, \apjs, 48, 95

\bibitem[Storey \& Hummer(1991)]{sto91} Storey, P.~J., \& Hummer, D.~G.\ 1991,
Computer Physics Communications, 66, 129

\bibitem[Verner et al.(1993)]{ver93} Verner, D.~A., Yakovlev, D.~G., Band, I.~M.,
\& Trzhaskovskaya, M.~B.\ 1993, Atomic Data and Nuclear Data Tables, 55, 233

\bibitem[Verner \& Ferland(1996a)]{ver96a} Verner, D.~A., \& Ferland, G.~J.\ 1996, \apjs, 103, 467

\bibitem[Verner et al.(1996b)]{ver96b} Verner, D.~A., Ferland, G.~J., Korista, K.~T.,
\& Yakovlev, D.~G.\ 1996, \apj, 465, 487

\bibitem[Zhang(1998)]{zha98} Zhang, H.~L.\ 1998, \pra, 57, 2640

\end{thebibliography}

\end{document}